\documentclass[aps,prx,reprint,superscriptaddress,nofootinbib]{revtex4-2}

\usepackage[utf8]{inputenc}
\usepackage{amsmath, siunitx}
\usepackage{amssymb}
\usepackage{physics}
\usepackage[dvipsnames]{xcolor}
\usepackage{xspace}
\usepackage{bbold}
\usepackage{graphicx}
\usepackage{dcolumn}
\usepackage{bm}
\usepackage{svg}
\usepackage[colorlinks=true]{hyperref}
\usepackage{cleveref}
\usepackage[autosize]{dot2texi}
\usepackage{tikz}
\usepackage{orcidlink}
\usepackage{mathtools}

\usepackage[caption=false,font=small]{subfig}

\definecolor{red}{HTML}{F25F5C}
\definecolor{blue}{HTML}{247BA0}
\definecolor{yellow}{HTML}{FFE066}
\definecolor{black}{HTML}{4F5051}
\definecolor{green}{HTML}{6FC1B3}
\definecolor{orange}{HTML}{F18F01}

\usetikzlibrary{decorations.markings,arrows.meta}
\tikzset
  {midarrow/.style={decoration={markings,mark=at position 0.5 with
     {\arrow[xshift=4pt]{Latex[length=6pt,#1]}}},postaction={decorate}}
  }

\usetikzlibrary{quantikz2,arrows.meta,positioning,fit,backgrounds}
\hypersetup{colorlinks=true, citecolor=blue}

\date{\today}

\begin{document}
\preprint{APS/123-QED}
\title{Coherent Quantum Evaluation of Collider Amplitudes for Effective Field Theory Constraints}

\author{Yacine Haddad${}^{\orcidlink{0000-0003-4916-7752}}$}
\email{yacine.haddad@cern.ch}
\affiliation{Albert Einstein Center for Fundamental Physics, Laboratory for High Energy Physics, University of Bern, Sidlerstrasse 5, CH-3012 Bern, Switzerland}
\affiliation{European Organization for Nuclear Research (CERN), Espl. des Particules 1211 Geneva 23, Switzerland}

\author{Kaidi Xu${}^{\orcidlink{0000-0003-2184-0829}}$}
\affiliation{Max Planck Institute of Quantum Optics, 85748 Garching, Germany}
\affiliation{Department of Physics and Arnold Sommerfeld Center for Theoretical Physics (ASC), Ludwig Maximilian University of Munich, 80333 Munich, Germany}
\affiliation{Munich Center for Quantum Science and Technology (MCQST), 80799 Munich, Germany}

\author{Vincent Croft${}^{\orcidlink{0000-0002-8731-4525}}$}
\affiliation{Applied Quantum Algorithms Leiden (aQa$^\text{L}$) and Leiden Institute of Advanced Computer Science (LIACS), Leiden University, 2333 CA Leiden, Netherlands}

\author{Jad C.~Halimeh${}^{\orcidlink{0000-0002-0659-7990}}$}
\affiliation{Department of Physics and Arnold Sommerfeld Center for Theoretical Physics (ASC), Ludwig Maximilian University of Munich, 80333 Munich, Germany}
\affiliation{Max Planck Institute of Quantum Optics, 85748 Garching, Germany}
\affiliation{Munich Center for Quantum Science and Technology (MCQST), 80799 Munich, Germany}
\affiliation{Department of Physics, College of Science, Kyung Hee University, Seoul 02447, Republic of Korea}

\author{Michele Grossi${}^{\orcidlink{0000-0003-1718-1314}}$}
\affiliation{European Organization for Nuclear Research (CERN), Espl. des Particules 1211 Geneva 23, Switzerland}

\date{\today}

\begin{abstract}
Precision measurements at electron–positron colliders provide stringent tests of the Standard Model and powerful probes of possible higher-dimensional interactions. We present a hybrid quantum–classical framework for computing leading-order helicity amplitudes for $e^+e^-\to \ell^+\ell^-$ scattering on gate-based quantum hardware and using the resulting cross sections to constrain both Standard Model couplings and effective field theory operators. In our approach, external kinematics are encoded into single-qubit Weyl spinors, and full helicity amplitudes are reconstructed by coherently combining diagrammatic contributions within a single quantum circuit. Classical post-processing yields physical amplitudes and differential cross sections that can be directly compared with collider data. As a proof of concept, we compute unpolarised angular distributions and perform binned likelihood fits to precision electron–positron measurements. The extracted bounds are statistically consistent with Standard Model expectations, demonstrating that quantum-assisted amplitude evaluation can interface directly with phenomenological analyses and experimental data. This work establishes a concrete pathway toward applying quantum computing to precision collider physics and effective field theory studies.


\end{abstract}
\maketitle

\section{Introduction}
\label{sec:intro}

Precision measurements at electron-positron colliders provide some of the cleanest tests of the electroweak interaction. Although the Large Electron-Positron collider (LEP) and the Stanford Linear Collider (SLC) shut down decades ago, their high-statistics measurements of total rates, angular distributions, and asymmetries at and above the $Z$ boson resonance remain benchmark datasets for constraining physics beyond the Standard Model (SM) \cite{Schael:2006pr, LEPEWWG:2013epw}. Effective field theories such as the SM Effective Field Theory (SMEFT) framework have emerged as the dominant model-agnostic approach for precision tests, and Beyond SM (BSM) searches \cite{Grzadkowski:2010es}, translating LEP/LHC observables into limits on Wilson coefficients via Monte-Carlo pipelines that evaluate matrix elements and integrate over phase space on equal footing. By expanding the standard model Lagrangian in powers of $(E/\Lambda)$ with dimension-six operators, EFTs systematically encode arbitrary new physics as measurable deviations from SM amplitudes. 


These classical pipelines face a fundamental bottleneck: the $O(N^2)$ scaling of interference-term calculations with the number of EFT parameters. Current global analyses constrain $\mathcal{O}(10^2)$ coefficients simultaneously, yet the complete basis contains thousands of independent operators. As future precision measurements at proposed facilities such as the FCC-ee will approach per-mil sensitivity to Higgs couplings and electroweak observables, broad parameter coverage becomes essential to avoid bias from overly restrictive assumptions, making the quadratic scaling of classical interference calculations a binding constraint on the reach of model-independent searches. Recent examples include SMEFT global-fit frameworks~\cite{giani_smefit_2023,Bartocci:2024MFVGlobalFit} and updates such as SMEFiT and \texttt{SMEFiT3.0}~\cite{celada_mapping_2024}, as well as simultaneous SMEFT-PDF analyses such as \texttt{SIMUnet}~\cite{costantini_simunet_2024}, which routinely fit $\mathcal{O}(10^2)$ effective parameters to hundreds of cross-section measurements across LEP and the Large Hadron Collider (LHC). 


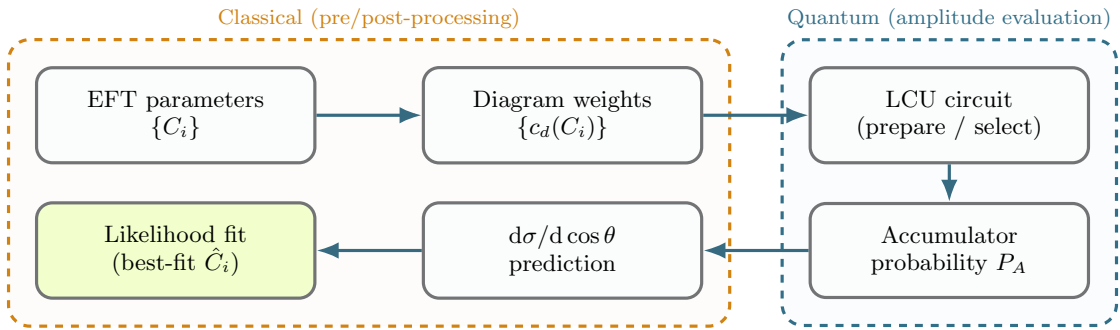
\begin{figure*}[ht!]
\centering
\begin{tikzpicture}[
  font=\small,
  node distance=5mm and 14mm,
  block/.style={
    rounded corners=6pt,
    draw=black!80,
    fill=black!2,
    very thick,
    minimum height=12mm,
    minimum width=25mm,     
    text width=25mm,        
    inner xsep=6mm,
    inner ysep=3mm,
    align=center,
  },
  fblock/.style={
    rounded corners=6pt,
    draw=black!80,
    fill=lime!20,
    very thick,
    minimum height=12mm,
    minimum width=25mm,     
    text width=25mm,        
    inner xsep=6mm,
    inner ysep=3mm,
    align=center,
  },
  arrow/.style={-Latex, very thick, draw=blue!60!black},
  qgroup/.style={
    draw=blue!70!black,
    dashed,
    very thick,
    rounded corners=8pt,
    inner sep=10pt,
    fill=blue!8,
    fill opacity=0.20,
    text opacity=1
  },
  cgroup/.style={
    draw=orange!80!black,
    dashed,
    very thick,
    rounded corners=8pt,
    inner sep=10pt,
    fill=orange!10,
    fill opacity=0.20,
    text opacity=1
  }
]

\node[block] (eft) {EFT parameters\\$\{C_i\}$};
\node[block, right=of eft] (wts) {Diagram weights\\$\{c_d(C_i)\}$};
\node[block, right=of wts] (lcu) {LCU circuit\\(prepare / select)};
\node[block, below=of lcu] (acc) {Accumulator\\probability $P_A$};
\node[block, left=of acc] (ds) {$\mathrm{d}\sigma/\mathrm{d}\cos\theta$\\prediction};
\node[fblock, left=of ds] (fit) {Likelihood fit\\(best-fit $\hat C_i$)};

\draw[arrow] (eft) -- (wts);
\draw[arrow] (wts) -- (lcu);
\draw[arrow] (lcu) -- (acc);
\draw[arrow] (acc) -- (ds);
\draw[arrow] (ds) -- (fit);

\begin{scope}[on background layer]
  \node[qgroup, fit=(lcu) (acc), label={[font=\footnotesize, text=blue!70!black]north:Quantum (amplitude evaluation)}] (Q) {};

  \node[cgroup, fit=(eft) (wts) (ds) (fit), label={[font=\footnotesize, text=orange!80!black]north:Classical (pre/post-processing)}] (C) {};
\end{scope}
\end{tikzpicture}
\caption{Algorithm overview: EFT parameters determine diagram coefficients $c_d$ and hence the LCU preparation weights. The quantum circuit evaluates the (helicity-resolved or helicity-averaged) amplitude via bracket-extraction blocks and a controlled accumulator rotation; classical post-processing converts the measured accumulator probability into $\mathrm{d}\sigma/\mathrm{d}\cos\theta$ and performs the binned likelihood fit to collider data.}
\label{fig:overview}
\end{figure*}

Gate-based quantum computers offer a complementary approach: complex phases and coherent interference are handled natively by unitary evolution, suggesting that \emph{amplitude formation} (the coherent combination of contributions) is the natural subtask to offload to quantum hardware. At the same time, phase-space integration and detector-level corrections remain classical. This observation underlies several recent proposals in high-energy physics (HEP), ranging from quantum parton showers \cite{Bauer:2019qps,Bauer:2024qps} to strategies for computing helicity amplitudes \cite{Bepari:2021}, quantum integration of decay rates\cite{qintegration_grossi25} and, more recently, direct quantum simulations of diagrammatic interferences \cite{Chawdhry:2025qcd}. In parallel, quantum-assisted EFT fits have been explored with annealing-based optimizers \cite{Criado:2023QFitter}, and variational circuits have been confronted with real collider data in the context of global parton-distribution (PDF) fits \cite{PerezSalinas:2021qPDF} and fragmentation function sampling\cite{Mart_nez_de_Lejarza_2025}. For broader context and benchmarks identified by the community, see the recent QC4HEP roadmap~\cite{di_meglio_quantum_2024} and related perspective reviews~\cite{afik_quantum_2025,funcke_review_2023,catterall_report_2022}, which survey quantum algorithms and early demonstrations across theory and experiment.

In this paper, we introduce a hybrid quantum algorithm that keeps the computational split explicit: we evaluate tree-level helicity amplitudes on a quantum device, while phase-space integration, flux/energy normalizations, and detector effects remain classical. Our quantum subroutine is built on a minimal two-component (Weyl) encoding, where each external massless fermion line is represented by a \emph{pair} of qubits (one undotted and one dotted) once the trivial $\sqrt{2E}$ normalization is factorized. Using standard spinor-helicity technology \cite{Elvang:2013Review, Dreiner:2010TwoComponent}, we implement compact expose-copy-unexpose blocks that extract angle and square brackets with Bell-basis maps, and we combine multiple tree-level diagrams, both SM and SMEFT contact terms, coherently via the Linear Combination of Unitaries (LCU) paradigm \cite{Chakraborty:2024LCU}. 

Complementary efforts pursue a more fully non-perturbative route based on quantum simulation of lattice gauge theories \cite{bauer_quantum_2023,di_meglio_quantum_2024,Halimeh2025ColdAtomQuantumSimulatorsGaugeTheories,halimeh2025quantumsimulationoutofequilibriumdynamics}. In the last few years, multiple platforms have reported proof-of-principle demonstrations of gauge-theory dynamics, ranging from HEP-relevant phenomena such as scattering and confinement to behavior related to quantum many-body physics, such as thermalization and its avoidance~\cite{dalmonte_lattice_2016, martinez_real-time_2016, yang_observation_2020, zhou_thermalization_2022, zhang_observation_2025, PhysRevD.103.094501, PhysRevD.105.074504, PhysRevLett.133.111901, PhysRevD.107.054513, PhysRevD.107.054512, PRXQuantum.5.020315, PhysRevD.109.114510, mildenberger_confinement_2025, schuhmacher2025observationhadronscatteringlattice, davoudi2025quantumcomputationhadronscattering, gyawali2025observationdisorderfreelocalizationusing, cochran_visualizing_2025, gonzalez-cuadra_observation_2025, crippa_analysis_2026, cobos2025realtimedynamics21dgauge}.

As a concrete testing ground, we target two canonical lepton-collider processes with clean helicity selection rules and precise historical measurements: dimuon production $e^+e^-\to\mu^+\mu^-$ and Bhabha scattering $e^+e^-\to e^+e^-$. We model electroweak effects as photon-plus-neutral-current contact/propagator terms in an EFT-inspired parametrization, and evaluate the corresponding helicity amplitudes as functions of the scattering angle. For Bhabha scattering, we coherently combine $s$- and $t$-channel contributions with ancillary qubits index registers (four LCU terms in the $\gamma+Z$ setup). For dimuons, the baseline $\gamma+Z$ setup uses two LCU terms in the $s$ channel. In both channels, helicity-label logic enforces numerator selection coherently, and diagram interference is generated within a single unitary circuit.

Our contributions are threefold. First, we provide a fully unitary circuit with explicit register-level mapping: two spinor qubits per external leg, four helicity-label qubits, three reversible Boolean ancillas, one accumulator, and an optional LCU index register. Secondly, we provide compact correctness arguments for Bell-inverse bracket extraction and the explicitly normalized LCU construction (with $\lambda=\sum_d|c_d|$), including the classical restoration of known prefactors. Finally, we demonstrate an end-to-end phenomenology loop in which quantum-generated angular predictions are fitted to legacy lepton-collider data in the $(\kappa_Z,g_V^2,g_A^2)$ basis, and we quantify finite-shot effects relevant for NISQ-style studies. The guiding principle is to use the quantum processor where coherent amplitude formation is structurally hardest classically, while keeping established HEP inference components classical. The overall workflow is summarized in Fig.~\ref{fig:overview}.

\section{Spinor-Helicity Amplitudes on Qubits}
\label{sec:spinor-qubit}

Massless fermions in four dimensions are described by two-component Weyl spinors that transform in the $(\tfrac{1}{2},0)$ (undotted, left-chiral) or $(0,\tfrac{1}{2})$ (dotted, right-chiral) representations of $\mathrm{SL}(2,\mathbb{C})$. Each two-component spinor is a normalized vector in a two-dimensional complex Hilbert space and can therefore be put in one-to-one correspondence with a single-qubit state~\cite{Bepari:2021}. Accordingly, our circuits allocate two spinor qubits per external fermion leg, one for the undotted spinor $\lambda$ and one for the dotted spinor $\tilde\lambda$. This observation enables us to directly translate the kinematic data of external, massless particles into gate sequences on a quantum device, following the standard two-component formalism described in~\cite{Dreiner:2010TwoComponent,Elvang:2013Review}.

For a particle with three-momentum
$\mathbf{p} = E(\sin\theta\cos\phi,\,\sin\theta\sin\phi,\,\cos\theta)$, a convenient phase convention for the associated Weyl spinors is
\begin{equation}
    \lambda_\alpha = \sqrt{2E} 
    \begin{pmatrix}
        \cos\frac{\theta}{2} \\
        e^{i\varphi} \sin\frac{\theta}{2}
    \end{pmatrix},
    \tilde\lambda_{\dot\alpha} = \sqrt{2E} 
    \begin{pmatrix}
        -e^{-i\varphi}\sin\frac{\theta}{2} \\
         \cos\frac{\theta}{2}
    \end{pmatrix}
    \label{eq:weyl-spinors}
\end{equation}
up to an arbitrary little-group phase\footnote{Any overall phase on $\lambda$ and $\tilde\lambda$ drops out at physical observables and is fixed here for convenience~\cite{Dixon:1996Amplitudes}.}. Stripping the trivial $\sqrt{2E}$ factor, the undotted spinor expression matches the canonical qubit state on the Bloch sphere,
\begin{equation}
    \ket{\psi(\theta,\phi)}
    = \cos\left(\tfrac{\theta}{2}\right)\ket{0}
    + e^{i\varphi}\sin\left(\tfrac{\theta}{2}\right)\ket{1}.
    \label{eq:qubit-bloch}
\end{equation}
Hence, each Weyl spinor can be encoded in a single qubit. In our implementation we always prepare \emph{both} the undotted and dotted momentum spinors for each external leg $i$, using two spinor qubits $(q_{\lambda_i},q_{\tilde\lambda_i})$,
\begin{equation}
\begin{split}
|\lambda_i\rangle &= U_3(\theta_i,\varphi_i,-\varphi_i)\,|0\rangle,\\
|\tilde\lambda_i\rangle &= U_3(\theta_i,-\varphi_i,\varphi_i)\,|0\rangle,
\end{split}
\end{equation}
where $U_3$ is Euler-angle single-qubit preparation~\cite{qiskit-Javadi-Abhari:2024kbf}, with $(\theta_i,\varphi_i)$ fixed by the external kinematics. External \emph{helicity} information is stored in a separate label register and used to coherently route the correct spinor qubits into the bracket-extraction blocks, keeping the overall circuit unitary and compatible with LCU-based diagram summation.

Spinor-helicity angle and square brackets are defined in the $\mathrm{SL}(2,\mathbb C)$ group as antisymmetric contractions
\begin{equation}
    \langle i\,j\rangle=\epsilon^{\alpha\beta}\lambda_{i,\alpha}\lambda_{j,\beta},\qquad
[i\,j]=\epsilon^{\dot\alpha\dot\beta}\tilde\lambda_{i,\dot\alpha}\tilde\lambda_{j,\dot\beta}.
\end{equation}

In our circuit, these antisymmetric contractions are accessed by projecting the two-spinor register onto the singlet state using a Bell-inverse map (App.~\ref{app:circuit-blocks}). These contractions encode the helicity structure of massless amplitudes compactly~\cite{Elvang:2013Review,Dreiner:2010TwoComponent}. In our encoding, these brackets are extracted coherently with a Bell-basis inverse. We define
\begin{equation}
  B^\dagger=(H\otimes I)\,\mathrm{CNOT},
\end{equation}
and for qubits $|a\rangle,|b\rangle$ carrying two spinors, a straightforward computational-basis expansion gives
\begin{equation}
   \langle11|B^\dagger|a\rangle\otimes|b\rangle =\frac{a_0 b_1-a_1 b_0}{\sqrt2}
   =\frac{\langle a b\rangle}{\sqrt2}
   \label{eq:bell_braket}
\end{equation}
and analogously for dotted spinors to obtain $[ab]/\sqrt{2}$. The $\ket{11}$ amplitude of the Bell-inverse pair is therefore the desired antisymmetric contraction, up to the fixed $1/\sqrt{2}$ factor. Two such blocks, one undotted and one dotted, expose products $\langle i k\rangle [j\ell]$, which by the two-spinor Fierz identity \cite{Dreiner:2010TwoComponent}
\begin{equation}
    \langle i|\sigma^\mu|j] \langle k|\sigma_\mu|\ell]=2 \langle i k\rangle [j\ell]
    \label{eq:Fierz-2spinor}
\end{equation}
reproduce the standard numerator algebra of $s$- or $t$-channel vector
exchange. 

A single multi-controlled $R_y$ rotation on an accumulator qubit copies the exposed bracket product onto the accumulator~\cite{PhysRevA.52.3457}. Importantly, no separate ``phase-copy'' $R_z$ step is required: the complex phase is inherited from the computational amplitude of the selected work-register branch (e.g., the $|11\rangle$ component after the Bell-inverse). In contrast, relative diagram phases are implemented either as diagonal phases on the LCU index register or as signs in the controlled-rotation angles. We then uncompute the Bell blocks, yielding a measurement-free, unitary estimate of the tree-level helicity amplitude on the accumulator, up to a known global energy normalization.

It is essential to note that the distinction between angle and square brackets corresponds to whether the contraction involves undotted or dotted Weyl spinors. While a single qubit can represent either $|\lambda\rangle$ or $|\tilde\lambda\rangle$ (up to conjugation), in our circuit we allocate \emph{two} qubits per external leg and prepare both $|\lambda_i\rangle$ and $|\tilde\lambda_i\rangle$ explicitly. This makes the subsequent bracket-extraction blocks uniform and keeps the overall construction fully unitary (no mid-circuit measurement), which is convenient for coherent diagram summation via LCU.

A complete derivation of the Weyl-qubit mapping and the Bell-based extraction of helicity brackets is given in App.~\ref{app:weyl-qubit} and App.~\ref{app:circuit-blocks}.

\section{Electroweak as an Effective Field Theory}
\label{sec:eft-setup}
In this work, we use the electroweak (EW) sector as an effective extension of QED. The basic idea is to start from photon-mediated QED amplitudes for $e^+e^-\to f\bar f$ and augment them with contact interactions that encode the effect of $Z$ exchange and, more generally, possible new physics in the neutral current~\cite{PhysRevLett.50.811}. At the level of four-fermion processes, this is conveniently described by an effective Lagrangian of the form,
\begin{equation}
  \mathcal{L}_\mathrm{eff}
  =
  \mathcal{L}_\mathrm{QED}
  +
  \sum_{f,f^\prime}
  \sum_{i,j=L,R}
  \frac{c_{ij}^{ff^\prime}}{\Lambda^2}
  \left(\bar f_i\gamma_\mu f_i\right)
  \left(\bar f_j^\prime\gamma^\mu f_j^\prime\right),
  \label{eq:leff-4f-EW}
\end{equation}
where $f,f^\prime\in\{e\mu\}$ label charged-lepton flavors, $i,j$ denote left- or right-handed chiralities, and the $c_{ij}^{ff^\prime}$ are dimensionless Wilson coefficients~\cite{PhysRevLett.50.811,Falkowski:2017pss}. In the Standard Model, these coefficients are fixed combinations of the usual neutral-current couplings $g_L^f,g_R^f$ and the $Z$ propagator, so that Eq.~\eqref{eq:leff-4f-EW} reproduces the tree-level $Z$ exchange when expanded around $s\ll m_Z^2$ in a Fermi-theory spirit \cite{DawsonEWPOfourF,IsidoriSMEFT}. Beyond the SM, the same coefficients can capture additional short-distance contributions as in SMEFT \cite{Brivio:2017vri}.

Concretely, for a neutral-current interaction
\begin{equation}
  \mathcal{L}_{Zff}
  =
  \frac{g}{\cos\theta_W}\,
  Z_\mu\,
  \sum_f
  \bar f \gamma^\mu
  \left(g_L^f P_L + g_R^f P_R\right) f,
\end{equation}
integrating out the $Z$ at tree level generates, to leading order in
$s/m_Z^2$,
\begin{equation}
  c_{ij}^{ff^\prime}
  \;\propto\;
  g_i^f\,g_j^{f^\prime},
  \qquad
  i,j\in\{L,R\},
\end{equation}
up to an overall normalization proportional to $m_Z^2/g^2$ and the propagator factor. In the SM, these relations are fixed, and the $c_{ij}^{ff^\prime}$ are not free parameters. In our numerical analysis, we instead treat them as effective Wilson coefficients to be extracted from data, using the quantum circuit to evaluate the corresponding helicity amplitudes. In the SM limit, the best-fit $c_{ij}^{ff^\prime}$ can be mapped back to the underlying $g_L^f,g_R^f$; in the presence of new physics, they should be interpreted as effective couplings encoding both SM and beyond-SM effects.

The structure in Eq.~\eqref{eq:leff-4f-EW} can be embedded in the Standard Model Effective Field Theory (SMEFT), where the SM Lagrangian is supplemented by all dimension-six operators consistent with the SM gauge symmetries~\cite{Brivio:2017vri,Grzadkowski:2010es}. In the Warsaw basis, the leptonic four-fermion operators relevant for $e^+e^-\to\mu^+\mu^-$ and Bhabha scattering are

\begin{align}
  \mathcal{O}_{\ell\ell}
  &=
  (\bar\ell\gamma_\mu \ell)\,
  (\bar\ell\gamma^\mu \ell),
  &
  \mathcal{O}^{(3)}_{\ell\ell}
  &=
  (\bar\ell\gamma_\mu \tau^I \ell)\,
  (\bar\ell\gamma^\mu \tau^I \ell)
  \nonumber\\[4pt]
  \mathcal{O}_{ee}
  &=
  (\bar e\gamma_\mu e)\,
  (\bar e\gamma^\mu e),
  &
  \mathcal{O}_{\ell e}
  &=
  (\bar\ell\gamma_\mu \ell)\,
  (\bar e\gamma^\mu e)
  \label{eq:4f-operators-Warsaw}
\end{align}
with $\ell$ the left-handed lepton doublet, $e$ the right-handed charged-lepton singlet and $\tau^I$ Pauli matrices in weak isospin space. After electroweak symmetry breaking and rotation to the mass basis, these operators generate exactly the kind of four-lepton contact terms appearing in Eq.~\eqref{eq:leff-4f-EW}; the $c_{ij}^{ff^\prime}$ are specific linear combinations of the Warsaw coefficients $C_k$ for four-lepton operators ~\cite{IsidoriSMEFT,Silva4F,DawsonEWPOfourF}. In this sense, our $e^+e^-\to\ell^+\ell^-$ fits can be viewed both as a determination of effective neutral-current couplings and as a restricted SMEFT fit in the four-lepton sector.

In the kinematic regime we consider, the leading SMEFT effect in the observables is the interference between the SM amplitudes and the dimension-six operators. Contributions quadratic in the Wilson coefficients are suppressed by additional powers of $s/\Lambda^2$ and are neglected here, in line with standard practice in SMEFT fits to electroweak and dilepton data at comparable energies ~\cite{Brivio:2017vri,IsidoriSMEFT}. The same helicity-based representation and circuit construction can be extended to Higgs- and gauge-sector SMEFT operators, allowing one, in principle, to treat the Higgs sector as an effective field theory and extract its couplings from data within the same quantum-in-the-loop framework.

\subsection{Benchmark processes at lepton colliders}
\label{subsec:processes}

We focus on two benchmark $e^+e^-$ processes with clean helicity structure and precise measurements that are easily verifiable theoretically: the Bhabha scattering $e^+e^-\to e^+e^-$ and the dimuon $e^+e^-\to\mu^+\mu^-$ production. At tree level in the SM, Bhabha scattering receives contributions from an $s$-channel annihilation diagram and a $t$-channel scattering diagram, both mediated by the photon (and by the $Z$ boson once electroweak effects are included. The dimuon channel proceeds via $s$-channel $\gamma/Z$ exchange only. The corresponding matrix elements are standard, and we list their helicity-decomposed expressions in App.~\ref{app:amplitudes}.

The four-lepton operators in Eq.~\eqref{eq:4f-operators-Warsaw} generate additional contact contributions to these amplitudes, which interfere with the SM in specific helicity channels. In our numerical analysis, we concentrate on the vector operators that respect the chiral structure of the SM neutral current, so that the SMEFT deformations appear as a simple rescaling of the corresponding helicity amplitudes~\cite{IsidoriSMEFT,Silva4F}. This choice allows us to encode the entire hard-scattering dependence on the Wilson coefficients in a small number of spinor--helicity numerators, which can be implemented efficiently using the circuits described in Sec.~\ref{sec:circuit}.

Explicit tree-level helicity amplitudes for $e^+e^-\to\mu^+\mu^-$ and Bhabha scattering, both in the SM and including the contact operators, are given in App.~\ref{app:amplitudes}.

\section{From amplitudes to circuit}
\label{sec:circuit}

Our implementation of the helicity amplitude uses a fixed register layout. For $N_{\rm diag}$ coherently summed terms, the LCU index register has $n_{\rm idx}=\lceil \log_2 N_{\rm diag}\rceil$ qubits, and the full circuit uses $n_{\rm idx}+16$ qubits. The 16 non-index wires are organized as follows: 8 spinor wires (two per external leg: undotted and dotted), 4 helicity-label wires $(h_1,h_2,h_3,h_4)$, 3 reversible Boolean ancillas $(v_1,v_2,b)$, and a single accumulator wire $A$.

For each external massless leg $i$, we prepare
\begin{align}
|\lambda_i\rangle &= U_3(\theta_i,\varphi_i,-\varphi_i)\,|0\rangle,\\
|\tilde\lambda_i\rangle &= U_3(\theta_i,-\varphi_i,\varphi_i)\,|0\rangle,
\end{align}
where $(\theta_i,\varphi_i)$ are the polar and azimuthal angles of the three-momentum of particle $i$ in the center-of-mass frame. In addition, a single helicity-label register is prepared either in a fixed computational basis state (for a chosen helicity configuration) or in superposition (to sum helicities coherently). Two Boolean ancillas implement the XOR/XNOR logic of the helicity conservation at each vertex, activating the appropriate numerator structures. The remaining registers are used for spinor-bracket extraction, diagram selection, and the final accumulator qubit.

\subsection{Helicity registers and coherent selection logic}
\label{subsec:helicity-register}

In addition to the spinor qubits, we allocate a four-qubit \emph{helicity-label} register $|h_1 h_2 h_3 h_4\rangle$ that stores the helicities of the external legs (incoming $e^-e^+$ and outgoing $\ell^-\ell^+$) in the computational basis. This register is either initialized to a fixed bit string (to evaluate a specific helicity channel) or, if no helicity configuration is specified, prepared in the uniform superposition $|+\rangle^{\otimes 4}$ via Hadamards. Because different helicity strings occupy orthogonal label states, measuring only the accumulator (and tracing over the helicity register) returns the average over helicity channels, which is the quantity needed for the unpolarised cross section.

To enforce the selection rules of the spinor-helicity numerators \emph{coherently} and without mid-circuit measurement, we compute a small set of Boolean predicates into ancilla qubits using reversible CNOT networks (XOR/XNOR logic). In the reference implementation, three ancillas $(v_1,v_2,b)$ are derived from the helicity bits and subsequently uncomputed; they act only as controls for the numerator blocks, ensuring that the overall circuit remains fully unitary and therefore compatible with LCU.

\subsection{Numerator selection within a diagram}
\label{subsec:numerator-selection}

For several diagrams (notably the $ffVff$ topologies relevant to $e^+e^-\to\ell^+\ell^-$), the spinor numerator can be written in more than one equivalent contraction pattern, with the appropriate form selected by the external helicities. We implement this with a single Boolean selector ancilla $b$: when $b=1$ we activate a \emph{Type~A} numerator block (e.g. $\langle 23\rangle[14]$), and when $b=0$ we activate a \emph{Type~B} block (e.g.\ $\langle 24\rangle[13]$).  Each numerator block consists of two Bell-inverse sandwiches that expose the required brackets, followed by a multi-controlled $R_y$ rotation that writes the corresponding complex amplitude onto the accumulator. The inactive numerator path is skipped by control logic, and all Boolean ancillas are uncomputed at the end of the block.

To implement a given helicity amplitude in spinor-helicity form, we must evaluate products of angle and square brackets, such as $\langle 24\rangle[13]$ for $e^+e^-\to\mu^+\mu^-$ in QED. As shown in Sec.~\ref{sec:spinor-qubit} and App.~\ref{app:circuit-blocks}, the Bell map $B$ provides an efficient way of doing so: after preparing the external qubits in the desired Bloch states, we apply $B^\dagger$ to a chosen pair of $\lambda$ or $\tilde\lambda$ qubits and read off the bracket from the amplitude of $|11\rangle$, Eq.~\eqref{eq:bell_braket}. Repeating this procedure for the pairs appearing in the numerator yields all required brackets as computational amplitudes that can be copied to an accumulator qubit through controlled $R_y$ blocks. The strength of the $R_y$ angle can be tuned depending on the chiral structure of the coupling considered. For instance, in the case of Z boson exchange, the two $R_y$ rotations in each diagram block can differ, while in the $\gamma$ exchange, they remain the same.

In practice, the circuit building blocks consist of (i) state-preparation rotations, (ii) Bell-inverse sandwiches for bracket extraction, and (iii) controlled phase and rotation gates that collect the diagram numerators onto the accumulator. A detailed gate-level implementation of these blocks is given in App.~\ref{app:circuit-blocks}.

\begin{figure*}[!ht]
    \centering
    \includegraphics[width=0.8\linewidth]{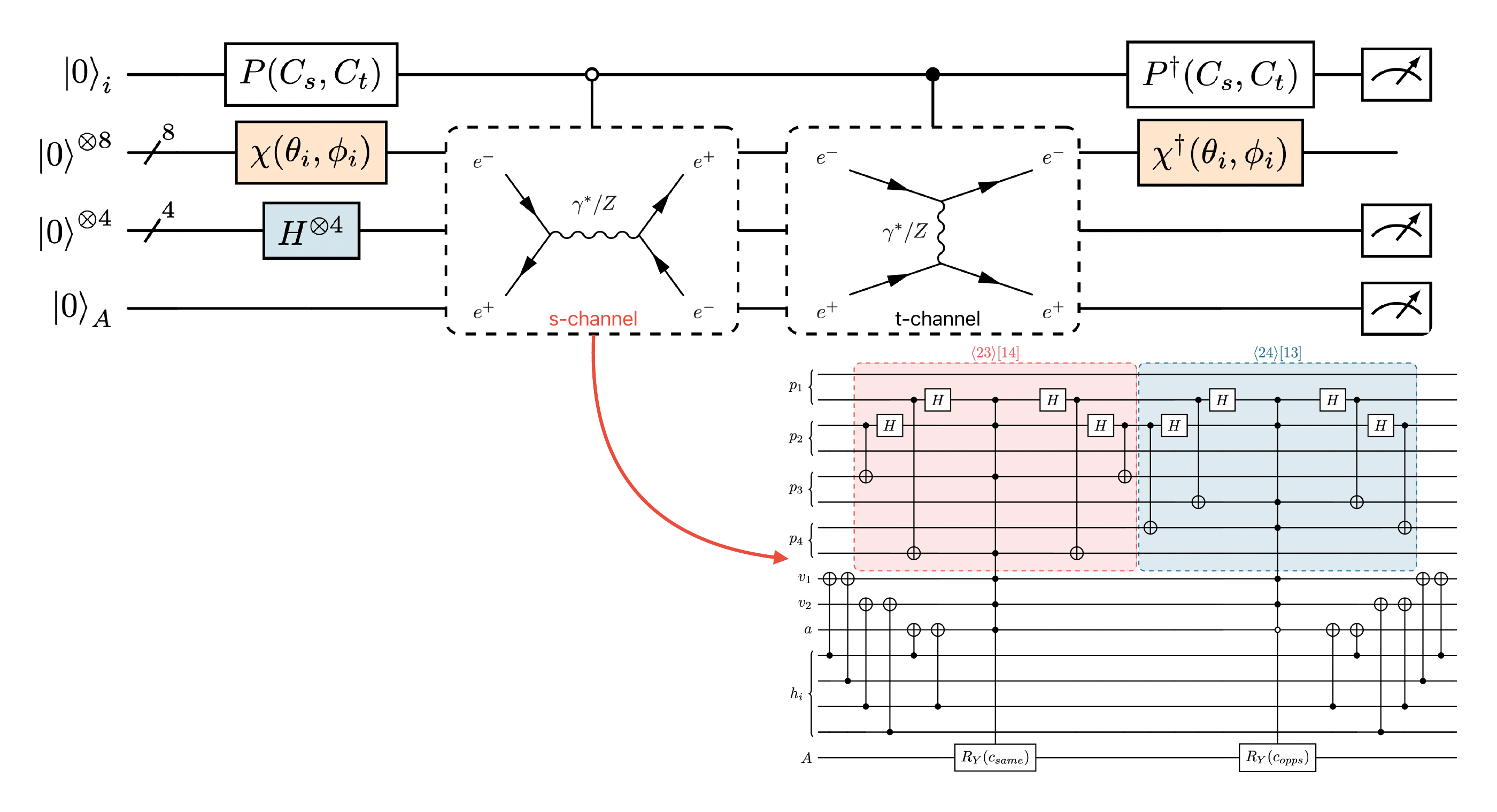}
    \caption{Compact quantum-circuit schematic for the $s$-channel block (full two-diagram layout shown above, with a zoom-in of the bracket-product subroutine at right). Each external leg $i$ is encoded as a two-wire Weyl bundle $(\lambda_i,\tilde\lambda_i)$, prepared by $W_i\equiv\chi(\theta_i,\phi_i)$ and unprepared by $W_i^\dagger=\chi^\dagger(\theta_i,\phi_i)$. The helicity logic block $\mathcal{F}_{\mathrm{hel}}$ coherently computes the flags $(v_1,v_2)$ and a selector bit $b$ that chooses the required spinor contraction. Bracket products are exposed with Bell-inverse gadgets $B^\dagger$: the \emph{Type A} path implements $\langle23\rangle[14]$ (red region, $b=1$), while the \emph{Type B} path implements $\langle24\rangle[13]$ (blue region, $b=0$). The resulting scalar is copied to the accumulator register $A$ via controlled $R_y$ rotations (with the appropriate coefficient, e.g.\ $c_{\mathrm{same}}$ or $c_{\mathrm{opps}}$), and then all work registers are uncomputed with $B$. For multi-diagram runs, the LCU index register is prepared with $P(C_s,C_t)$ and unprepared with $P^\dagger(C_s,C_t)$.}
    \label{fig:circuit}
\end{figure*}

To coherently sum the contributions of several Feynman diagrams, we employ a linear-combination-of-unitaries (LCU) construction.\cite{ChildsWiebe:2012LCU,Berry:2015TruncatedTaylor,Chakraborty:2024LCU}  We introduce a small \emph{diagram register} of $n_d$ index qubits, whose computational basis states $|d\rangle$ label the different diagrams (for the processes considered here $n_d\leq 2$). A preparation unitary $V$ creates a superposition
\begin{equation}
|D\rangle
= V|0\cdots 0\rangle
= \sum_{d}\,\omega_d\,|d\rangle,
\end{equation}
with complex weights $\omega_d$ chosen to match the desired coupling and propagator factors of each diagram. Conditioned on the state of the diagram register, we then apply controlled versions of the circuit blocks that expose and copy the spinor-bracket numerators onto an \emph{accumulator} qubit, as sketched in Fig.~\ref{fig:circuit}.

After uncomputing the diagram register (by applying $V^\dagger$), we post-select the index register back onto $|0\cdots 0\rangle$. On this successful branch, the accumulator amplitude is proportional to the coherently summed scattering amplitude,

\begin{equation}
\mathcal{A}_{\text{QC}} \;\propto\; \lambda^{-1}\sum_d c_d\,\mathcal{M}_d,\qquad \lambda=\sum_d |c_d|\,,
\end{equation}
so that the corresponding probability estimates $|\mathcal{A}_{\text{QC}}|^2$ are rescaled by the known factor $\lambda^2$ in classical post-processing. This procedure captures diagram interference exactly within a single unitary circuit layer.

In principle, one could estimate the inner product with a SWAP test; this approach requires extra ancillas and returns moduli rather than complex amplitudes. The Bell-inverse method directly extracts the signed, complex spinor contractions and maintains a circuit measurement-free (unitary-only) approach, which is essential for LCU-based coherent summation.

For $e^+e^-\to\mu^+\mu^-$, the base circuit uses 16 logical qubits: eight for the external spinor registers (two per external leg), four helicity-label qubits, three Boolean ancillas for helicity and numerator selection, and one accumulator qubit. When coherently summing multiple diagrams with LCU, an additional index register of $n_\text{idx}=\lceil\log_2 N_\text{diag}\rceil$ qubits is required (e.g.\ $n_\text{idx}=1$ for combining the $s$- and $t$-channel Bhabha diagrams).

\begin{figure}
    \centering
    \includegraphics[width=0.8\linewidth]{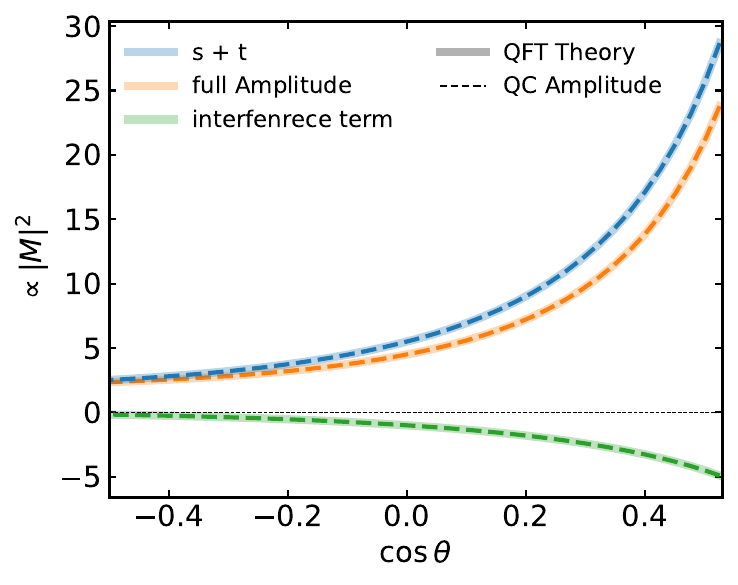}
    \caption{Closure test comparing $|\mathcal{M}_{\rm QC}|^2$ to an independent classical spinor-helicity evaluation as a function of $\cos\theta$. Examples include the dimuon $s$-channel and Bhabha $s/t$ contributions, the interference term, and their LCU combination. }
    \label{fig:lcu_amplitudes}
\end{figure}

\section{Fit to collider data}
\label{sec:fit}

\subsection{Experimental input}
\label{subsec:data}
For proof of concept, we confront our quantum-computed amplitudes with measurements of the differential cross sections for $e^+e^-\to\mu^+\mu^-$ and Bhabha scattering from PEP and LEP. At $\sqrt{s}=29~\mathrm{GeV}$ we take the single-differential $d\sigma/d\cos\theta$ distribution measured by the MAC collaboration using the MAC detector at the PEP storage ring, \cite{ash_precision_1985,derrick_new_1986} binned in the polar angle $\theta$ of the outgoing lepton in the centre-of-mass frame. We use the tabulated values and quoted uncertainties as archived in the open-access \texttt{HEPData} repository~\cite{hepdata:Maguire:2017ypu,hepdata.3246.v1/t1,hepdata.3835.v1/t1,hepdata.6636.v1/t1}.

In each case, we denote by $n_i$ the experimental value in bin $i$, either a cross section or an event count normalized to the integrated luminosity, with an uncertainty $\sigma_i$ that combines the quoted statistical and systematic errors in quadrature. The corresponding theory prediction $\mu_i(\vec{c})$ is obtained from the quantum amplitudes as described below.

\subsection{From quantum amplitudes to binned predictions}
\label{subsec:predictions}

For a fixed parameter point $\vec c$, scattering angle $\theta$, and center-of-mass energy $\sqrt{s}$, we choose the LCU preparation weights to reproduce the propagator and coupling prefactors of the contributing diagrams. In the $\gamma+Z$ setup this gives

\begin{equation}
\vec C_{\rm LCU}^{\rm bh}=\left(\frac{\alpha}{s},\,\frac{\alpha}{t},\,g_s,\,g_t\right),
\qquad
\vec C_{\rm LCU}^{\mu\mu}=\left(\frac{\alpha}{s},\,g_s\right),
\end{equation}
for Bhabha scattering and dimuon production, respectively.\footnote{In practice, we rescale the LCU coefficient vector by a common kinematic factor, $\tilde c_d \equiv s\,c_d$, to keep the coefficients $\mathcal{O}(1)$ and reduce their dynamic range. Since the LCU state preparation depends only on the normalized weights $|c_d|/\lambda$ with $\lambda=\sum_d|c_d|$, this global rescaling does not change the circuit probabilities. The corresponding overall factor is restored in the classical normalization used to convert the reconstructed $|\mathcal M|^2$ into differential cross-sections, so the final predictions are unchanged.} Here $\alpha\equiv\alpha_{\rm em}$ is the QED fine-structure constant, $s$ and $t$ are the Mandelstam invariants, and $g_{s,t}$ denotes the effective neutral-current coupling entering the $s$- and $t$-channel blocks. In the contact-interaction (EFT) treatment with $\Lambda=m_Z$, the leading neutral-current contribution is encoded as $g_{s,t}\propto -\,\kappa_Z\,\frac{G_F}{\Lambda^2}$.

Each circuit evaluation returns samples from the joint distribution of the LCU index register $I$, the accumulator qubit $A$, and the helicity-label register $H\equiv(h_1h_2h_3h_4)$. Since we work with unpolarised observables, we trace over the helicity labels and define the acceptance probability on the successful LCU branch,
\begin{equation}
P_{\rm acc}(1;\theta,\vec c) = \sum_{h_1h_2h_3h_4}P(I=0, A=1, h_1h_2h_3h_4;\theta,\vec c).
\end{equation}
Post-selection on the LCU index register yields an amplitude proportional to $\lambda^{-1}\sum_d C_d\,\mathcal M_d$, with the known normalization $\lambda=\sum_d |C_d|$ restored in classical post-processing. In our register convention, tracing over the coherently prepared helicity-label superposition introduces a fixed overall normalization, so that the reduced squared matrix element is reconstructed as

\begin{equation}
|\mathcal M(\theta,\vec c)|^2
= 32~\lambda^2\,P_{\rm acc}(1;\theta,\vec c),
\end{equation}
where the factor $32$ is a kinematics-independent normalization fixed by the circuit conventions. In particular, tracing over a helicity-label register prepared as $|+\rangle^{\otimes 4}$ yields a uniform helicity average and introduces an overall factor $2^{-4}$ at the probability level. The remaining constant factor arises from the fixed normalization of the Bell-projection used in the circuit.

The reconstructed $|\mathcal M|^2$ is converted to the experimental differential observables as
\begin{equation}
\frac{d\sigma}{d\cos\theta} = \frac{\pi\alpha_{\rm em}^2}{s}~|\mathcal M|^2,
\label{eq:dsigdcos}
\end{equation}
including the standard $\mathrm{GeV}^{-2}\!\to\mathrm{nb}$ conversion factor.

To compare with data, we integrate Eq.~\eqref{eq:dsigdcos} over the experimental bin ranges in $\cos\theta$, applying the same acceptance cuts used in the early dimuon and Bhabha analyses~\cite{ash_precision_1985,derrick_new_1986,fernandez_electroweak_1987}. This yields the binned predictions, $\mu_i(\vec c)$, that enter the likelihood. In our unpolarised setup, the same-helicity and opposite-helicity contributions are collected into the two weights $(c_{\rm same},c_{\rm opp})$ in both channels.

\subsection{Likelihood and parameter inference}
\label{subsec:likelihood}

We perform a binned likelihood fit to the measured differential cross sections using the parameter set
\begin{equation}
\vec{\theta} = (\kappa_Z, g_V^2, g_A^2),
\end{equation}
where $\kappa_Z$ is an overall rescaling of the $Z$ exchange contribution and $(g_V^2,g_A^2)$ are the squared vector and axial couplings entering the unpolarised Bhabha amplitudes. 

Since we coherently sum all the helicity configurations, the total cross section of the circuit output can be treated as unpolarised Bhabha scattering with negligible fermion masses. The observable cross section is largely insensitive to the \emph{signs} of $(g_V,g_A)$ and depends dominantly on the combinations
\begin{align}
c_{\rm RL,LR} & =\frac14(g_V^2+g_A^2),\\
c_{\rm RR,LL} & = \frac14(g_V^2-g_A^2),
\label{eq:csamecopp}
\end{align}
which are the helicity-weight combinations entering the $Ry(c)$ accumulators in our quantum-circuit construction~\cite{LEPEWWG:2013epw,PDG:2024RPP}. We therefore fit directly in $(g_V^2,g_A^2)$ to remove the discrete sign degeneracy and improve convergence of the numerical minimization. 

This parametrization is motivated by the fact that in contact-interaction (EFT) interpretation, unpolarised annihilation and Bhabha scattering at high energy are primarily sensitive to two independent helicity-conserving combinations. In particular, the LL and RR structures are not distinguishable in unpolarised Bhabha angular distributions. Similarly, one typically identifies LR and RL in the absence of polarization information at the reconstruction level. We therefore impose the simplifying relations $c_{LL}=c_{RR}$ and $c_{LR}=c_{RL}$ reducing the fit to two independent helicity combinations (equivalently $c_{\rm same}$ and $c_{\rm opp}$) plus the $Z$-rescaling parameter $\kappa_Z$. This assumption is justified for our proof-of-concept setup and matches the known insensitivity of Bhabha data to separating LL from RR in unpolarised fits~\cite{derrick_new_1986, fernandez_electroweak_1987}. 

\begin{figure}[!h]
\centering
\includegraphics[width=0.8\linewidth]{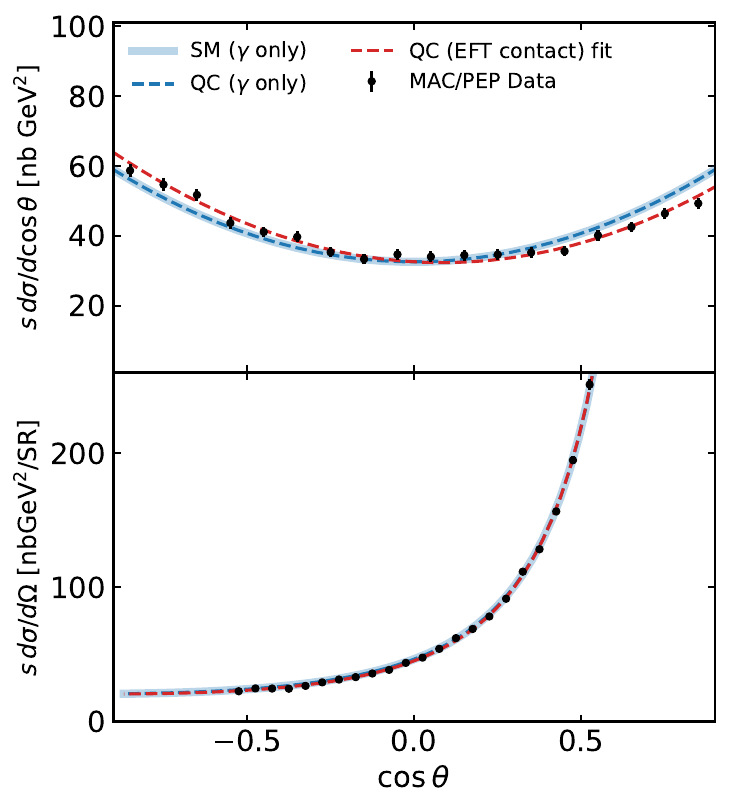}
\caption{Differential cross-section overlays with data. Data points with error bars are compared to the SM prediction
and representative EFT benchmark curves derived from the quantum-computed amplitudes.}
\label{fig:dsigma-data}
\end{figure}

Now, let $n_i$ denote the measured value in bin $i$ (cross section or luminosity-normalized count) with quoted uncertainty $\sigma_{i,\rm exp}$. For parameters $\vec c=(g_V^2,g_A^2)$, the theory prediction $\mu_i(\vec c)$ is computed from the quantum-evaluated matrix element combined with the standard flux and phase-space factors. When the circuit is evaluated with a finite number of shots, $\mu_i$ is replaced by a stochastic estimator $\hat\mu_i$ with an additional sampling variance $\sigma^2_{i,\rm shot}(\vec c)$. Following the standard treatment of finite simulation statistics in likelihood fits using the Barlow-Beeston approach and its subsequent formulations~\cite{barlow_fitting_1993,Conway2011,Arguelles2019}, we incorporate this by augmenting the per-bin variance and minimizing the corresponding negative log likelihood,
\begin{equation}
\begin{split}
& -2\ln\mathcal{L}_{\rm shot}(\vec c)
= \\
& \sum_{i=1}^M
\left[
\frac{\left(n_i-\hat\mu_i(\vec c)\right)^2}{\sigma_{i,\rm exp}^2+\sigma_{i,\rm shot}^2(\vec c)}
+\ln\left(\sigma_{i,\rm exp}^2+\sigma_{i,\rm shot}^2(\vec c)\right)
\right]
\end{split}
\label{eq:nll}
\end{equation}
up to additive constants. In the infinite-shot limit $\hat\mu_i(\vec\theta)\to \mu_i(\vec\theta)$ and $\sigma_{i,\rm shot}(\vec\theta)\to0$, which reduces Eq.~\eqref{eq:nll} to the usual analytic Gaussian likelihood. The construction of $\sigma_{i,\rm s}$, the shot-scaling study, and the extraction of parameter uncertainties using profile likelihood are detailed in App.~\ref{app:fit-details} and App.~\ref{app:shots}.

\section{Results and outlook}
\label{sec:results}

A central validation of our workflow is that likelihood surfaces reconstructed from quantum measurement outcomes reproduce the analytic likelihood derived from the underlying scattering amplitude. In our setting, the quantum circuit provides an estimator of the event probabilities as a function of the EFT parameters, which we then feed into a classical negative log-likelihood. The fit itself remains classical, while the quantum device provides the scattering amplitudes and cross-sections as functions of the scattering angles. Parameter inference is performed using standard maximum-likelihood machinery. Figure~\ref{fig:dsigma-data} summarizes the differential cross-section comparison between data and our quantum-generated theory predictions, illustrating both the Standard Model baseline and representative EFT benchmark points.

\begin{figure}[!ht]
\centering
\includegraphics[width=0.8\linewidth]{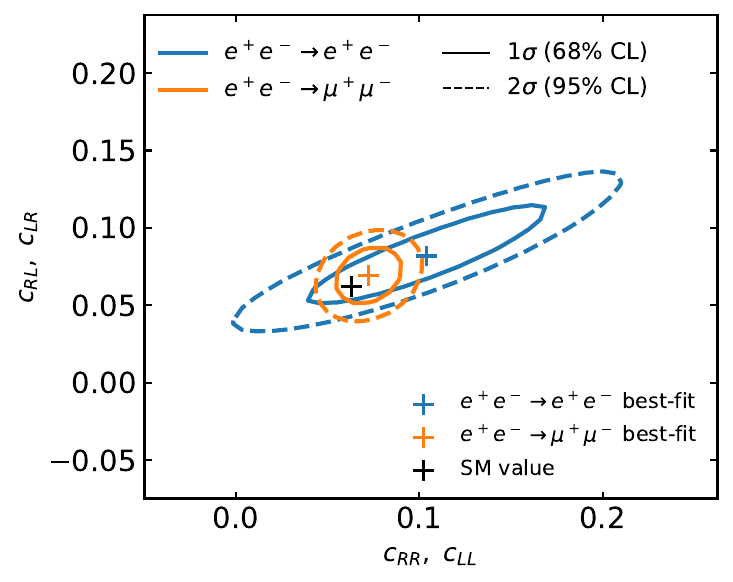}
\caption{ Likelihood contours in the $(c_{RR/LL},c_{RL/LR})$ plane extracted from the quantum-evaluated likelihood, shown at $68\%$ and $95\%$ confidence level (solid and dashed, respectively). The best-fit point from the fit is indicated along with the SM reference point. This figure illustrates the correlation structure induced by interference terms and provides a direct test that the circuit-encoded amplitude reproduces the analytic likelihood geometry.}
\label{fig:contours_wilson}
\end{figure}

We report best-fit values and uncertainties in the $c_{\rm same}$ and $c_{\rm opp}$ basis (see Figure~\ref{fig:contours_wilson}), which is directly connected to the helicity-weight combinations. These could be mapped to the chiral coupling $g_A$ and $g_V$ as defined in Eq.~\eqref{eq:csamecopp}. We find that the parameters are compatible with the electroweak (EW) contribution at the $1\sigma$ level and are consistent with the results from ~\cite{fernandez_electroweak_1987}. 

\begin{figure}[t]
  \centering
  \includegraphics[width=0.8\linewidth]{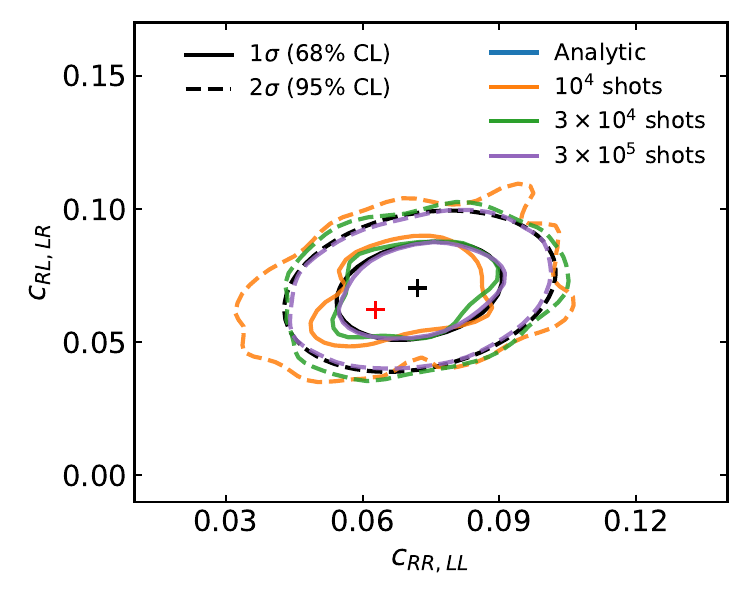}
  \caption{Comparison of analytic likelihood contours to shot-based likelihood reconstructions at increasing shot budgets for the $e^+e^-\to\mu^+\mu^-$ channel. As the number of shots increases, the reconstructed confidence regions converge toward the analytic reference, consistent with an unbiased estimator limited by finite sampling.}
  \label{fig:contours_wilson_shots}
\end{figure}

We also compare the likelihood geometry in two-dimensional parameter planes. Figure~\ref{fig:contours_wilson_shots} shows the $(c_{RR/LL},c_{RL/LR})$ constraints obtained from shot-based likelihood evaluations at increasing shot budgets, overlaid with the analytic reference. The reconstructed confidence regions converge toward the analytic contours as the number of shots increases, and already at $\mathcal{O}(10^4)$ shots, the agreement is nearly complete. Importantly, the \emph{orientation} and \emph{curvature} of the contours match the analytic Fisher-information geometry, demonstrating that the circuit correctly captures both the dominant terms and the interference structure responsible for parameter correlations.

\begin{figure}[!t]
  \includegraphics[width=0.9\linewidth]{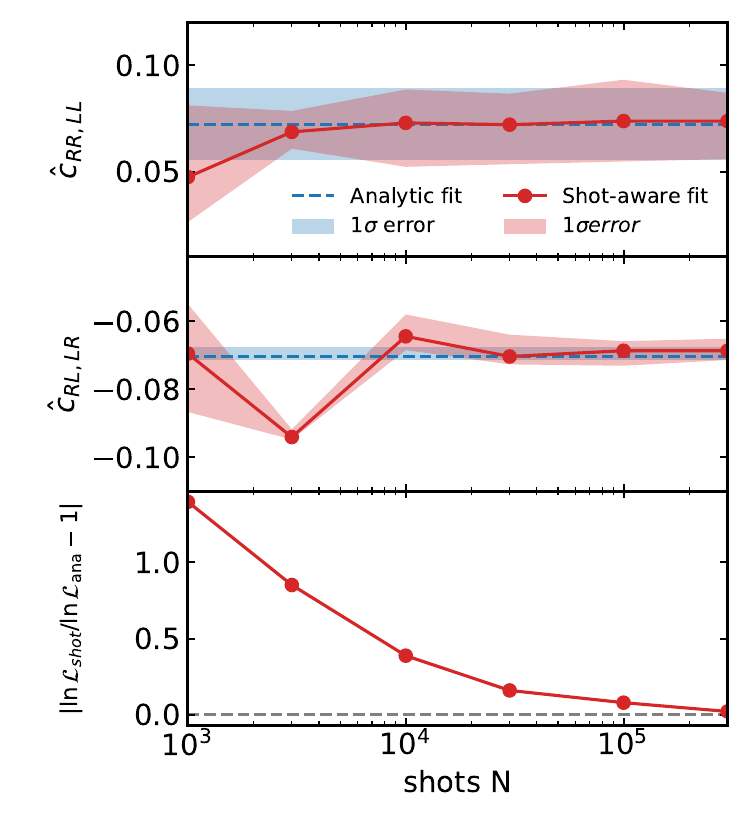}
  \centering
  \caption{Convergence of the finite-shot inference with the number of circuit shots $N$ (log scale). Top and middle panels show the best fitted coefficients $\hat c_{RR,LL}$ and $\hat c_{RL,LR}$ as a function of number of shots: the orange dashed line (with orange band) denotes the analytic/infinite-statistics fit and its $1\sigma$ uncertainty, while the blue markers/line (with blue band) show the corresponding shot-aware fit obtained from circuit samples. Bottom panel: absolute fractional deviation of the shot-aware log-likelihood from the analytic one, $\bigl|\ln \mathcal{L}_{\rm shot}/\ln \mathcal{L}_{\rm ana}-1\bigr|$, illustrating the goodness of fit approaching the analytic limit as $N$ increases.}
  \label{fig:wilks_shot_scaling}
\end{figure}

To quantify finite-shot effects on inference outputs, we study the shot dependence of the best-fit point and the $1\sigma$ interval extracted from $\Delta\mathrm{NLL}$ using Wilks' theorem. Figure~\ref{fig:wilks_shot_scaling} summarizes these results for $c_{RR,LL}$ and $g_{RL,LR}$. The best-fit values fluctuate around the analytic best fit, with error bars obtained from the shot-based runs. At low shot counts ($\lesssim 10^4$), the fluctuations are visibly larger, consistent with multinomial shot noise in the estimated probabilities; as shot counts increase, the best-fit values stabilize, and the inferred intervals approach the analytic reference.

A key point is that the observed deviations do not exhibit a coherent drift with shot number: within uncertainties, they are consistent with \emph{zero-bias} sampling fluctuations. This supports the interpretation that the quantum circuit realizes an unbiased likelihood estimator and that the dominant limitation is finite sampling rather than a systematic error in amplitude encoding.

Beyond parameter values, we test convergence at the level of the likelihood optimum and the predicted distributions. Figure~\ref{fig:wilks_shot_scaling} also shows that the goodness of fit, defined as the difference relative to the analytic optimum, decreases with increasing shots, and the prediction-level discrepancy decreases accordingly. Together, these diagnostics confirm that the inference converges not only in parameter space but also in the space of physical predictions.

\subsection{Outlook}
\label{subsec:outlook}

The circuit-based amplitude computation demonstrated here addresses a practical bottleneck in precision inference. In many EFT-style analyses, predictions for large parameter sets require combining a growing number of interfering contributions of order $O(N^2)$. In our construction, interference is generated coherently within a single unitary, and the addition of further diagram terms corresponds to extending the LCU block rather than explicitly enumerating all interference structures, thereby avoiding costly preprocessing steps. This makes the approach naturally aligned with next-generation lepton-collider programs (e.g., FCC-ee), where clean final states, high statistics, and beam polarization can be exploited to isolate helicity channels.

The quantum circuit presented here implements an evaluation oracle for the squared amplitude $|\mathcal{M}(\theta; c)|^2$ as a function of scattering angle $\theta$ and coupling parameters $c$. In the language of computational learning theory, this constitutes a generative learning problem: given $N$ scattering events sampled from the true distribution $|\mathcal{M}(\theta; g_\mathrm{true})|^2$, the task is to infer the parameters $c_\mathrm{true}$. The LCU-based Feynman circuit returns the probability density at arbitrary phase space points for any choice of parameters. This access model is naturally stronger than sample-only access, in which one can merely draw events from a candidate distribution but cannot evaluate its density directly.

For effective field theories at weak coupling, individual amplitude evaluations remain classically tractable via perturbation theory, placing the present construction outside the BQP-complete regime for strongly-coupled or high-precision scattering calculations. The circuit nonetheless instantiates the formal structure required for quantum learning separations: a quantum process generates the training data, and a quantum device evaluates the model. Established results show that computational separations between quantum and classical learners can arise even when sample complexity is equivalent, provided the learning problem possesses exploitable structure under quantum query access\cite{Gyurik:2023quj}. The present work thus provides theoretical infrastructure for extensions into regimes where perturbative shortcuts fail, and guaranteed quantum advantages emerge.

At the same time, several limitations remain. The present circuits are deep for near-term demonstrations, and we provide a \emph{resource-estimation and methodology} deep dive in (App.~\ref{app:hardware}). More fundamentally, scaling to higher multiplicity ($2\to N$) processes and to higher perturbative orders (NLO) requires further algorithmic development, since phase-space integration and (at hadron colliders) PDF convolutions introduce additional classical costs. A promising direction is to exploit the fully parametric nature of the quantum amplitude model to reduce or replace Monte-Carlo reweighting in EFT inference, while keeping the integration task classical.

Finally, improving the pedagogical presentation of the eigen-/Bell-basis extraction steps (App.~\ref{app:circuit-blocks}) and strengthening the derivation of the LCU prefactor (App.~\ref{app:lcu-closure}) will help connect the implementation details to standard QFT intuition.

\bigskip
\footnotesize
\begin{acknowledgments}
    Y.H. is supported by the Open Quantum Institute (OQI); the OQI itself is an initiative hosted by CERN, born at GESDA, and supported by UBS. M.G.~and Y.H. are supported by the CERN Quantum Technology Initiative. K.X.~and J.C.H.~acknowledge funding by the Max Planck Society, the Deutsche Forschungsgemeinschaft (DFG, German Research Foundation) under Germany excellence Strategy - EXC-2111 - 390814868, and the European Research Council (ERC) under the European Horizon Europe research and innovation program (Grant Agreement No.~101165667)-ERC Starting Grant QuSiGauge. Views and opinions expressed are, however, those of the author(s) only and do not necessarily reflect those of the European Union or the European Research Council Executive Agency. Neither the European Union nor the granting authority can be held responsible for them. This work is part of the Quantum Computing for High-Energy Physics (QC4HEP) working group.
\end{acknowledgments}
\normalsize

\appendix
\section{Weyl spinors and qubit correspondence}
\label{app:weyl-qubit}

The free Dirac equation decouples in the massless limit into the Weyl equations, In the spinor-helicity formalism a massless four-momentum is represented as a bispinor $p_{\alpha\dot\alpha}=p_\mu\sigma^\mu_{\alpha\dot\alpha}=\lambda_\alpha\,\tilde\lambda_{\dot\alpha}$, where $\lambda$ (undotted) and $\tilde\lambda$ (dotted) are two-component Weyl spinors defined up to the little-group rescaling $\lambda\to t\,\lambda$, $\tilde\lambda\to t^{-1}\tilde\lambda$ with $t\in\mathbb{C}^\times$. Physical amplitudes are built from the antisymmetric contractions $\langle ij\rangle=\epsilon^{\alpha\beta}\lambda_{i,\alpha}\lambda_{j,\beta}$ and $[ij]=\epsilon^{\dot\alpha\dot\beta}\tilde\lambda_{i,\dot\alpha}\tilde\lambda_{j,\dot\beta}$.

For an external leg $i$ with momentum direction $(\theta_i,\varphi_i)$, we fix a convenient phase convention and strip the trivial $\sqrt{2E_i}$ normalization factor, which is reinstated in classical post-processing. Each (normalized) Weyl spinor then corresponds to a pure qubit state on the Bloch sphere. Crucially, our circuit always prepares \emph{both} spinors for each external fermion leg using two qubits,

\begin{align}
|\lambda_i\rangle &= U_3(\theta_i,\varphi_i,-\varphi_i)\,|0\rangle\\
|\tilde\lambda_i\rangle &= U_3(\theta_i,-\varphi_i,\varphi_i)\,|0\rangle
\end{align}
which is exactly the behavior of the state-preparation primitive used in our PennyLane implementation. Helicity information is stored separately in a label register (Sec.~\ref{subsec:helicity-register}) and only affects which spinor contractions are activated.

\section{Circuit building blocks}
\label{app:circuit-blocks}

In this appendix, we collect the explicit circuit constructions used in Sec.~\ref{sec:circuit}. We first show how to prepare the external Weyl spinors as single-qubit Bloch states, and then prove that a simple Bell-inverse sandwich suffices to extract angle and square brackets as computational-basis amplitudes, cf.~Eq.~\eqref{eq:bell_braket}.

For a particle with momentum direction $\hat{\mathbf p}=(\sin\theta\cos\phi,\sin\theta\sin\phi,\cos\theta)$, the helicity eigenstates are the two-component Weyl spinors defined in Eq~\ref{eq:weyl-spinors}. These are precisely the pure qubit states on the Bloch sphere obtained by
\begin{align}
 \chi_{-}\equiv|LH\rangle &= R_z(\phi)R_y(\theta)R_z(-\phi)|0\rangle \\
 \chi_{+}\equiv|RL\rangle &= R_z(\phi)R_y(\theta)R_z(-\phi)|1\rangle
\end{align}
which are the solutions to the Weyl equation for massless particles $(\boldsymbol{\sigma}\cdot\hat{\mathbf p})\chi_{\pm}=\pm \chi_{\pm}$. Direct substitution yields the textbook identity, which is the standard equivalence between the Bloch sphere and the Pauli vector.

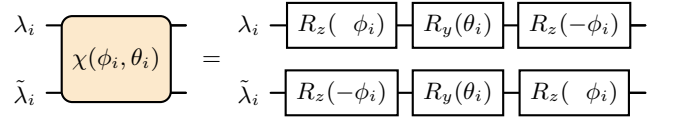
\begin{figure}[!ht]
\centering
\begin{quantikz}[column sep=0.2cm, row sep=0.3cm]
\lstick{$\lambda_i$}      & \gate[2, style={rounded corners,fill=orange!20, inner ysep=-5pt}]{\chi(\phi_i,\theta_i)} & \qw\\
\lstick{$\tilde\lambda_i$}&     & \qw
\end{quantikz}
=
\begin{quantikz}[column sep=0.2cm, row sep=0.3cm]
\lstick{$\lambda_i$}      & \gate{R_z(~~\phi_i)}\hphantom{} & \gate{R_y(\theta_i)} & \gate{R_z(-\phi_i)} &\qw \\
\lstick{$\tilde\lambda_i$}& \gate{R_z(-\phi_i)} & \gate{R_y(\theta_i)} & \gate{R_z(~~\phi_i)}\hphantom{} &\qw
\end{quantikz}
\caption{Weyl spinors state preparation called for each external fermionic leg. }
\label{app:fig:state-prep}
\end{figure}

Two qubits are used per external leg $i$: one to hold the undotted spinor $|\lambda_i\rangle$ (angle spinor) and one for the dotted $|\tilde\lambda_i\rangle$ (square spinor). For massless kinematics, one can choose phases so that
\begin{equation}
\tilde\lambda_i = \lambda_i^* \quad\Rightarrow\quad
|\tilde\lambda_i\rangle = R_z(-\phi_i)R_y(\theta_i) |h_i\rangle
\end{equation}
is implementable unitarily and refrains from using anti-unitary conjugation. The sketch of the Weyl spinor state preparation is shown in Figure~\ref{app:fig:state-prep}.

The helicity spinors normalization which they scale as $\lambda_i=\sqrt{2E_i}\chi_{h_i}$ and $\tilde\lambda_i=\sqrt{2E_i}\tilde\chi_{h_i}$, are done by multiplying by $\prod \sqrt{2E}$ factor in a post-processing step.

In the helicity amplitude formalism, the angle and square brackets are estimated by contracting the two Weyl spinors via the $SU(2)$ antisymmetric operator, 
\begin{equation}
    \langle i j\rangle \equiv \lambda_i^T\varepsilon\lambda_j, \qquad 
    [i j]\equiv \tilde\lambda_i^T\varepsilon\tilde\lambda_j, \qquad
    \varepsilon = i\sigma_Y
\end{equation}
If we consider two qubits $|a\rangle=a_0|0\rangle+a_1|1\rangle$ and $|b\rangle=b_0|0\rangle+b_1|1\rangle$, then

\begin{equation}
\langle \psi^-|a\otimes b\rangle =\frac{a_0 b_1 - a_1 b_0}{\sqrt{2}} =\frac{a^T\varepsilon b}{\sqrt{2}},    
\end{equation}
where $|\psi^-\rangle=(|01\rangle-|10\rangle)/\sqrt{2}$ is the singlet. 
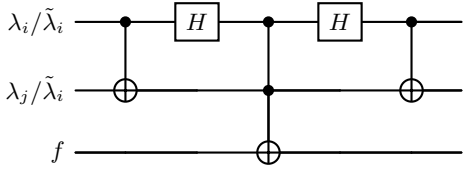
\begin{figure}[!ht]
\centering
\begin{quantikz}
\lstick{$\lambda_i/\tilde\lambda_i$} & \ctrl{1} & \gate{H} & \ctrl{2} & \gate{H} & \ctrl{1} & \qw \\
\lstick{$\lambda_j/\tilde\lambda_i$} & \targ{}  & \qw      & \ctrl{1} & \qw      & \targ{}  & \qw \\
\lstick{$f$}         & \qw      & \qw      & \targ{}  & \qw      & \qw      & \qw
\end{quantikz}
\caption{Angle bracket (square) $\langle i j\rangle$ ($[ij]$) on $\lambda_i,\lambda_j$ ($\tilde\lambda_i,\tilde\lambda_j$) using the $B^\dagger$ operator defined in Eq~\ref{eq:bell_braket}.}
\end{figure}
This is easily provable by expanding in the computational basis. The unitary $B=\mathrm{CNOT}\cdot(H\otimes I)$ maps $|11\rangle\mapsto|\psi^-\rangle$. Hence $B^\dagger|\psi^-\rangle=|11\rangle$. Subsequently, any $\lambda$ qubit pair $(i,j)$, the amplitude on $|11\rangle$ after the $\bra{11}B^\dagger\ket{i,j}=\frac{\langle ij\rangle}{\sqrt{2}}$ and similarly $[ij]/\sqrt{2}$ when applied to the $\tilde\lambda$ qubits. This provides you with the complex bracket value as a computational basis amplitude, not just its magnitude.

\section{Tree-level helicity amplitudes scattering for dimuon and Bhabha processes}
\label{app:amplitudes}

We consider the process $e^-(p_1)e^+(p_2)\to\mu^-(p_3)\mu^+(p_4)$ in the massless limit. At tree level in the SM, the amplitude receives contributions from $s$-channel photon and $Z$ exchange,

\begin{equation}
\mathcal{M}_\mathrm{SM}
= \mathcal{M}_\gamma + \mathcal{M}_Z,
\end{equation}
with
\begin{equation}
\mathcal{M}_\gamma
= -e^2\,\frac{Q_e Q_\mu}{s}~~
\bar v(p_2)\gamma^\mu u(p_1)~~
\bar u(p_3)\gamma_\mu v(p_4)
\end{equation}
and an analogous expression for $\mathcal{M}_Z$ in terms of the vector and axial couplings $g_{V,A}^f$.\cite{IsidoriSMEFT} In the spinor-helicity formalism, and for a particular helicity configuration, these currents reduce to products of angle and square brackets. For example, for the helicity assignment $(-+ -+)$ one finds
\begin{equation}
\mathcal{M}_\gamma(1^- 2^+ 3^- 4^+)
= -2e^2\,\frac{Q_e Q_\mu}{s}\
\langle 24\rangle[13]
\end{equation}
while other helicity configurations can be obtained by parity and crossing symmetry~\cite{Dixon:1996Amplitudes}. The four-lepton contact interactions in Eq.~\eqref{eq:leff-4f-EW} contribute additional $s$-channel amplitudes of the form

\begin{equation}
\mathcal{M}_\mathrm{EFT}
= \frac{1}{\Lambda^2}
\sum_{i,j=L,R}
c_{ij}^{e\mu}
\left(\bar e_i\gamma_\mu e_i\right)
\left(\bar\mu_j\gamma^\mu \mu_j\right),
\end{equation}
which translates into a simple rescaling of the relevant helicity amplitudes when written in spinor-helicity form. For instance, a purely right-handed operator contributes,
\begin{equation}
\mathcal{M}_{RR}(1^- 2^+ 3^- 4^+)
= \frac{c_{RR}^{e\mu}}{\Lambda^2}\,
\langle 24\rangle[13],
\end{equation}
up to an overall phase convention. In our circuit implementation, we encode the SM and SMEFT contributions by assigning appropriate weights to the Bell-extracted brackets $\langle 24\rangle$ and $[13]$. Summing over helicities and averaging over initial polarizations yields the unpolarised differential cross section quoted in Eq.~\eqref{eq:dsigdcos}.\cite{IsidoriSMEFT, Silva4F}

\subsection{Bhabha scattering $e^+e^- \rightarrow e^+e^-$}
\label{subapp:bhabha}

For Bhabha scattering we consider $e^-(p_1)e^+(p_2)\to e^-(p_3)e^+(p_4)$, again treating the electron as massless. The tree-level SM amplitude receives contributions from both an $s$-channel annihilation diagram and a $t$-channel scattering diagram,

\begin{equation}
\mathcal{M}_\mathrm{SM}
= \mathcal{M}_s + \mathcal{M}_t,
\end{equation}
with
\begin{align}
\mathcal{M}_s
&= -e^2\,\frac{Q_e^2}{s}\,
\bar v(p_2)\gamma^\mu u(p_1)\;
\bar u(p_3)\gamma_\mu v(p_4),
\\[4pt]
\mathcal{M}_t
&= -e^2\,\frac{Q_e^2}{t}\,
\bar u(p_3)\gamma^\mu u(p_1)\;
\bar v(p_2)\gamma_\mu v(p_4).
\end{align}
In spinor--helicity notation, these currents again reduce to products of angle and square brackets, with the $t$-channel amplitude related to the annihilation case by crossing.

The four-electron contact operators from Eq.~\eqref{eq:4f-operators-Warsaw} contribute additional $s$- and $t$-channel pieces that interfere with the SM. For a purely right-handed operator, one finds, for instance,
\begin{align}
\mathcal{M}_{RR,s}(1^- 2^+ 3^- 4^+)
&= \frac{c_{RR}^{ee}}{\Lambda^2}\,
\langle 24\rangle[13],
\\[4pt]
\mathcal{M}_{RR,t}(1^- 2^+ 3^- 4^+)
&= -\frac{c_{RR}^{ee}}{\Lambda^2}\,
\langle 23\rangle[14],
\end{align}
where the minus sign in the $t$-channel reflects the Fermi statistics of the identical fermions, in our circuit, we implement these contributions by selecting the appropriate Bell-extracted brackets for the pairs $(2,4)$, $(1,3)$, $(2,3)$, and $(1,4)$, and combining them with diagram-specific weights via the LCU block described in Sec.~\ref{sec:circuit}.

\section{Fit procedure and shot-statistics scaling}
\label{app:fit-details}

\subsection{Likelihood model and analytic limit}
\label{app:likelihood-model}

Equation~\eqref{eq:nll} defines the $-2\log$ likelihood used in the fit. The key point is that at a finite shot count, the quantum prediction is itself a random estimator, so the likelihood must account for both experimental measurement uncertainty and prediction (sampling) uncertainty. This situation is directly analogous to likelihood fits with finite Monte Carlo statistics, for which the Barlow-Beeston construction and later nuisance-parameter formulations provide a principled treatment~\cite{barlow_fitting_1993,Conway2011}.

In the Gaussian (large-sample) limit, profiling or marginalizing the auxiliary per-bin prediction nuisance parameters yields an effective likelihood in which the bin variance is augmented by the prediction variance, leading to Eq.~\eqref{eq:nll}; related likelihoods for stochastic predictions and their statistical properties are discussed in Ref.~\cite{Arguelles2019}.

To make the analytic limit explicit, note that for each bin $i$ the circuit returns a shot estimator $\hat\mu_i(\vec\theta)$ with $\mathrm{Var}(\hat\mu_i)=\sigma^2_{i,\rm shot}(\vec\theta)$. As $N_{\rm shots}\to\infty$ (or equivalently as the effective shot budget increases), $\hat\mu_i(\vec\theta)\to \mu_i(\vec\theta)$ and $\sigma_{i,\rm shot}(\vec\theta)\to0$, so the likelihood reduces to the standard analytic Gaussian likelihood used in conventional electroweak fits~\cite{PDG:2024RPP}.

\subsection{Number of shots and error propagation}
\label{app:shots}

In our implementation, the binned prediction is an affine rescaling of shot-estimated probabilities, $\hat\mu_i(\vec\theta)=\sum_d a_{id}(\vec\theta)\,\hat p_d$ that includes known LCU normalizations and phase-space weights. Therefore, the finite-shot variance propagates to the observable as

\begin{equation}
\sigma^2_{i,\rm shot}(\vec\theta)
\equiv \mathrm{Var}[\hat\mu_i(\vec\theta)]
\simeq
\sum_d \left(\frac{\partial \mu_i}{\partial p_d}\right)^2 \mathrm{Var}(\hat p_d)
\end{equation}
and 
\begin{equation}
    \mathrm{Var}(\hat p_d)=\frac{p_d(1-p_d)}{S_d},
\end{equation}
where in practice an estimate $p_d\to\hat p_d$ is used. This yields the expected scaling $\sigma_{i,\rm shot}\propto S_d^{-1/2}$ and is the quantity entering Eq.~\eqref{eq:nll}. Because $\sigma^2_{i,\rm shot}(\vec\theta)$ depends on the fit parameters, the logarithmic term in Eq.~\eqref{eq:nll} is required for a proper likelihood up to constants.

\subsection{Uncertainties from profile likelihood}
\label{app:profile-likelihood}

At finite shot count, the objective in Eq.~\eqref{eq:nll} is a noisy estimator because $\hat\mu_i(\vec c)$ fluctuates, which can render local curvature (Hessian) error estimates unstable. We therefore extract parameter uncertainties from the profile likelihood, following the standard MINOS approach~\cite{JamesRoos1975}. In this method, for a parameter of interest $\alpha$ and nuisance parameters $\nu$, define the profile likelihood ratio
\begin{equation}
\lambda(\alpha)=\frac{\mathcal{L}_{\rm shot}(\alpha,\hat\nu(\alpha))}{\mathcal{L}_{\rm shot}(\hat\alpha,\hat\nu)} ~~\to ~~
q(\alpha)\equiv -2\ln\lambda(\alpha)
\end{equation}
Under the regularity conditions and in the asymptotic regime, $q(\psi)$ is distributed as $\chi^2_k$ with $k$ the number of parameters of interest (Wilks' theorem) \cite{Cowan2011,wilks_large-sample_1938}. Thus, for one parameter of interest, approximate confidence intervals are obtained from $q(\psi)\le 1.0$ (68.27\% CL) and $q(\psi)\le 4.0$ (95.45\% CL); for two parameters of interest jointly, the corresponding contours are $q\le 2.30$ (68.27\%) and $q\le 6.18$ (95.45\%). These are the standard HEP thresholds used to quote likelihood-profiled uncertainties.

In our implementation, we compute $q$ using two-dimensional scans in $(g_V^2,g_A^2)$ (with $\kappa_Z$ fixed to its best-fit value) and obtain one-dimensional profile curves by minimizing along the orthogonal direction on the scan grid. Finite-shot fluctuations can introduce small-scale roughness in the scanned surface; we optionally apply a mild Gaussian smoothing to the $\Delta(-2\ln\mathcal{L})$ grid and verify that the extracted intervals are stable with respect to variations in the smoothing strength and improve systematically with increasing shot count. We note that Wilks' asymptotic approximation can fail for small effective sample sizes, boundary solutions, or non-regular models; see Ref.~\cite{Algeri2019} for a detailed discussion and possible remedies.

\section{LCU implementation, interference, and closure tests}
\label{app:lcu-closure}

Here we show that the LCU block used in Sec.~\ref{sec:circuit} indeed implements a coherent sum of diagram amplitudes, including all interference terms, on the accumulator qubit. Consider a set of diagrams labeled by $d=1,\dots,D$, each associated with a unitary $U_d$ that prepares the corresponding diagram numerator on the accumulator and work registers,

\begin{equation}
U_d\,|0\rangle_{\rm acc}\,|0\rangle_{\rm work}
=
\left(
\beta_d\,|0\rangle_{\rm acc}
+ \gamma_d\,|1\rangle_{\rm acc}
\right)
|\psi_d\rangle_{\rm work}.
\label{eq:lcu-Ud}
\end{equation}
The complex coefficient $\gamma_d$ is proportional to the helicity amplitude $\mathcal{M}_d$ for diagram $d$ (up to an overall normalization common to all diagrams)~\cite{ChildsWiebe:2012LCU}.

We introduce an $n_d$-qubit diagram register initialized in a superposition,
\begin{equation}
|D\rangle
= V|0\cdots 0\rangle
= \sum_{d}\,\omega_d\,|d\rangle,
\end{equation}
where $V$ is a preparation unitary and the weights $\omega_d$ encode the coupling and propagator factors associated with each diagram. We then apply the controlled unitary,

\begin{equation}
\mathcal{U}
= \sum_d |d\rangle\langle d|
\otimes U_d,
\end{equation}
acting on the combined system $|D\rangle\otimes|0\rangle_{\rm acc}\otimes|0\rangle_{\rm work}$. This yields
\begin{equation}
\mathcal{U}\,
|D\rangle|0\rangle_{\rm acc}|0\rangle_{\rm work}
=
\sum_d \omega_d\,|d\rangle
\left(
\beta_d\,|0\rangle_{\rm acc}
+ \gamma_d\,|1\rangle_{\rm acc}
\right)
|\psi_d\rangle_{\rm work}.
\label{eq:lcu-after-U}
\end{equation}
Finally, we uncompute the diagram register by applying $V^\dagger$. Writing $V^\dagger|d\rangle = \sum_k v_{kd}|k\rangle$ and projecting onto the reference state $|0\cdots 0\rangle_{\rm diag}$, we obtain
\begin{align*}
\langle 0\cdots 0|_{\rm diag}&
V^\dagger
\mathcal{U}
|D\rangle|0\rangle_{\rm acc}|0\rangle_{\rm work}
 \\ 
&=\sum_d \omega_d v_{0d}
\left(
\beta_d |0\rangle_{\rm acc}
+ \gamma_d |1\rangle_{\rm acc}
\right) 
|\psi_d\rangle_{\rm work}
\nonumber\\
&\equiv
\mathcal{N}
\left(
\tilde\beta|0\rangle_{\rm acc}
+ \tilde\gamma|1\rangle_{\rm acc}
\right)
|\Psi\rangle_{\rm work},
\end{align*}
where $\mathcal{N}$ is an overall normalization and $\tilde\gamma \propto \sum_d \omega_d \gamma_d$ is the coherent sum of the diagram amplitudes. The probability of measuring the accumulator in $|1\rangle$ is therefore
\begin{equation}
P_{\rm acc}(1)
= |\tilde\gamma|^2
\propto
\Big|\sum_d \omega_d \gamma_d\Big|^2,
\end{equation}
which contains all interference terms between different diagrams. Since the same overall proportionality factor appears for all processes and kinematic configurations, it can be absorbed into a known normalization of the amplitude, as done in Sec.~\ref{sec:fit}. This proves that our LCU block faithfully reproduces the coherent sum of diagram amplitudes in standard QFT~\cite{ChildsWiebe:2012LCU,Berry:2015TruncatedTaylor}.

\section{Implementation on Hardware}
\label{app:hardware}

The hardware implementation costs are heavily contingent on the decomposition of Toffoli gates and the specific connectivity between qubits. A detailed breakdown of these costs across various hardware layouts is provided in Table \ref{tab:cost}, where $N_{\text{1qb}}$, $N_{\text{2qb}}$, and $N_{\text{3qb}}$ denote the number of one-, two-, and three-qubit gates, respectively, and $D_{\text{2qb}}$ represents the two-qubit gate depth. We first evaluate a heavy-hex layout (\texttt{ibm-marrakech}), performing circuit decomposition and transpilation via \texttt{qiskit-ibm-runtime}. For the Bhabha channel, this results in 3,328 one-qubit gates and 1,021 two-qubit gates, with a two-qubit depth of 739. Consequently, the cumulative two-qubit depth across all four channels approaches 3,000. In contrast, we consider all-to-all connectivity both with and without a native \texttt{CCX} (Toffoli) gate. As shown in Table \ref{tab:cost}, all-to-all connectivity reduces the two-qubit depth to 635; by utilizing a native \texttt{CCX} gate, this depth is further minimized to 231 (at the cost of a 32 \texttt{CCX} gate depth). Circuit transpilation for the H2-2E was executed using the \texttt{pytket-quantinuum} package at optimization level 3. This process incorporates Pauli graph-based optimization methods and assumes \texttt{ZZPhase} as the native entangling gate.

\begin{table}[!h]
\begin{tabular}{l|cccccc}
\hline
Layout         & $N_{\rm 1qb}$ & $N_{\rm 2qb}$ & $D_{\rm 2qb}$ & $N_{\rm 3qb}$ & Backend    & opt level \\ \hline\hline
hyper-hex      & 3301          & 1023          & 712           &               & marrakech  & 3         \\ 
all-to-all     & 1136          & 811           & 635           &               & H2-2E      & 3         \\ 
all-to-all
w \texttt{CCX} & 311           & 264           & 231           & 32            & Aer        & 3         \\ \hline
\end{tabular}
\caption{The gate counts with respect to different geometries of hardware layout and native gate sets.}
\label{tab:cost}
\end{table}

\bibliography{bibliography}

\end{document}